# High energy-resolution transient ghost absorption spectroscopy

Alok Kumar Tripathi[1#], Yishai Klein[1#], Edward Strizhevsky[1], Flavio Capotondi[2], Dario De Angelis[2], Luca Giannessi[2,3], Matteo Pancaldi[2], Emanuele Pedersoli[2], Kevin C. Prince[2,4], Or Sefi[1], Young Yong Kim[5,6], Ivan A. Vartanyants[5] and Sharon Shwartz[1*]

[1] Physics Department and Institute of Nanotechnology and advanced Materials, Bar Ilan University, Ramat Gan, 52900, Israel
[2] Elettra-Sincrotrone Trieste, Strada Statale 14-km 163.5, Basovizza, 34149, Trieste, Italy
[3] Istituto Nazionale di Fisica Nucleare, Laboratori Nazionali di Frascati, Via E. Fermi 54, 00044 Frascati (Roma), Italy
[4] Department of Chemistry and Biotechnology, School of Science, Computing and Engineering Technology, Swinburne University of Technology, Melbourne, VIC, Australia
[5] Deutsches Elektronen-Synchrotron DESY, Notkestr. 85, 22607 Hamburg, Germany
[6] European XFEL, Holzkoppel 4, 22869 Schenefeld, Germany
# Contributed equally
*Corresponding author: sharon.shwartz@biu.ac.il

**We demonstrate the measurement of ultrafast dynamics using ghost spectroscopy and a pump-probe approach with an optical pump and a short-wavelength radiation probe. The ghost spectroscopy approach is used to overcome the challenge of the strong intensity and spectrum fluctuations at free-electron lasers and to provide high -spectral resolution, which enables the measurement of small energy shifts in the absorption spectrum. We exploit the high resolution to explore the dynamics of the charge carrier excitations and relaxations and their impact on the photoinduced structural changes in silicon by measuring the variation of the absorption spectrum of a Si(100) membrane near the silicon $L_{2,3}$ edge and the accompanying edge shifts in response to the optical illumination.**

Transient absorption spectroscopy is an important technique that is used for the study of ultrafast dynamics in several types of materials such as organic [1], inorganic [2], and biological [3] systems. When it is applied with X-rays or extreme-ultraviolet (XUV) radiation, this short wavelength spectroscopy probes the electronic structure of materials and provides element specific information on the charge and the spin structures as well as on the bonding configurations [4]. Transient absorption spectroscopy at short wavelengths is mainly performed with laboratory high harmonic generation (HHG) sources [1] or increasingly with free-electron lasers (FELs), since these machines provide intense pulses with durations ranging from attoseconds to hundreds of femtoseconds [5]. In most cases, the measurement is performed using the pump-probe approach in which a short pulse generated by the FEL probes the process triggered by a pump stimulus at various delays between the two pulses. The pump may be either an optical laser or the FEL itself [6].

However, since most FELs utilize the process of self-amplified spontaneous emission (SASE) for the generation of the ultrashort pulses, the pulse intensity and the spectrum suffer from significant fluctuations. In principle it is possible to overcome this challenge by measuring the spectra with two single-shot spectrometers, that are mounted one before and the second after the sample, but in practice this task is very challenging [7]. This is because the fluctuations are strong and the number of photons at the detectors is not always sufficient to overcome the shot noise, especially when the photons are detected by a large number of pixels at the spectrometer detector. For example, when high-energy resolution is required or when the signal is weak. In addition, the alignment of the spectrometer after the sample is highly sensitive to the angle between the beam that passes through the sample and the spectrometer, thus the alignment and calibration of the second spectrometer is very challenging and time consuming. Finally, the scheme can only be applied to transmissive samples, which strongly limits the choice of materials that can be studied.

A second commonly used approach is to employ narrow bandwidth radiation, which is obtained either by using a monochromator or by using a seeding scheme [8]. With this approach the absorption spectrum is reconstructed by scanning the photon energy of the input beam and by measuring the total intensity before and after the sample. The energy resolution of this method is determined by the bandwidth of the input beam, and thus high energy resolution requires narrow bandwidth radiation. When the measurement of the phenomena of interest requires the measurement of a broad spectrum with several experimental sampling points, there is a fundamental trade-off between the resolution and the measurement time. Especially, when monochromators are used there is an additional trade-off between the flux and the resolution since the narrower the monochromator bandwidth the lower the flux on the sample. In cases where the signal is weak or when high resolution is required, this trade-off can be a significant disadvantage of the monochromator approach. In a recent example, a monochromator was used to measure NEXAFS by detecting the electron yield of the sample [9].

A third approach for the measurement of absorption with FELs, which could overcome the challenges of the above methods is ghost spectroscopy (GS), a form of correlation spectroscopy [10, 11]. With this approach SASE radiation is used and the spectrum of the beam is measured before the sample by a single shot spectrometer but after the sample only the total transmitted or reflected intensity is measured. The spectrum is reconstructed by using a computational approach, which is based on similar computational algorithms that are used for the ghost imaging, which has been studied extensively [12] and in recent years demonstrated with X-rays [13, 14]. Since GS utilizes the spectral fluctuation of the beam, the method is very suitable for SASE radiation. Furthermore, since no spectrometer is required after the sample, the alignment of the detector after the sample is easy and the method can be used for weak signals that it would be challenging to measure with a spectrometer due to signal-to-noise (SNR) constraints. GS has been demonstrated recently with soft X-rays and with XUV radiation for the measurement of the static absorption spectrum and in one case the method has been used for the measurement of the dynamics induced by a FEL [15]. In [16], GS was benchmarked with respect to standard methods and showed much better spectral resolution than the SASE radiation and comparable resolution to the seeded scheme. The fact that high resolution with GS has been achieved with a measurement time that is comparable to the SASE measurement time suggests an avenue for lifting the trade-off between the resolution and the measurement time in standard approaches for spectroscopy. This is appealing for pump probe measurements since they require a large number of measurements. This letter demonstrates the advantages of the high energy-resolution transient GS method for the measurement of ultrafast dynamics. While we have demonstrated the high energy resolution in previous study [16], here we extend the method for the pump-probe approach measurements. As with the static measurements, the transient approach provides very high -energy resolution with a significantly reduced number of measurements.

The sample we explore in the present work is a thin Silicon membrane. We apply the transient GS method to provide important information on the simultaneous dynamics of electronic excitations and relaxations and photoinduced variation in the band structure of the sample. We exploit the very high resolution of our method (estimated based on our previous work to be 35 meV [16]) to distinguish between the various simultaneous phenomena. The high resolution is extremely important for thin films like our sample because of the existence of significant impurity states with narrow line width. Thanks to the high resolution of GS we can also resolve the details of the spectrum near the edges and distinguish between the contributions of the $L_2$ and $L_3$ edges (the separation between the edges is about 500 meV) [16]. We choose to explore the absorption spectrum in the energy range near the L-edge of silicon, which corresponds to excitation of the 2p core electrons to the valence and conduction bands, and thus can be used to explore the dynamics of the optically induced charge distribution.

The experiment was conducted at the DiProI beamline by using the FEL-2 light source at the FERMI FEL facility [8]. To demonstrate GS, FEL-2 was operated in SASE mode, with pulses at photon energies between 99 eV and 102 eV with step size of 250 meV for the measurements of the Silicon $L_2$ and $L_3$ edges. The experimental setup is presented in Fig. 1. The pulse duration of the SASE radiation is estimated to be about 250 fs.

The shot-by-shot photon spectra were measured with the Pulse-Resolved Energy Spectrometer Transparent and Online (PRESTO) [17], which is mounted at FERMI at the exit of the undulators and before beam transport to the end stations. A variable line spacing grating delivers most of the radiation to the end-stations, while the weaker first diffraction order is used to measure the source spectrum. At the DiProI end station, we mounted the sample in the direct beam and a photodiode was mounted immediately downstream of the sample to measure the total transmission through the sample. The optical excitations were triggered by a pump laser at 3.1 eV (400 nm) with a pulse duration of about 100 fs, beam dimensions of 700 × 670 μm² and fluence of 8.5 mJ/cm². In the experiment, we measured various delays from -5 ps to 150 ps and the angle between pump radiation and the sample was ∼ 5 deg. In order to cover the full energy range of the Si L-edge from 98.5 eV to 102 eV, the SASE central emission energy was set to 11 values each one with an average spectral width of 500 meV. At each of the SASE photon energies we measured 2000 shots, as we have demonstrated that this number of recorded shots is sufficient to obtain spectroscopic information [16].

To reconstruct the spectrum for each SASE central energy, we utilized the following reconstruction procedure: we represent the intensities of the N shots measured by the photodiode by a vector **T** (test data). The spectra of the shots are represented by the matrix **A** for which every row is the spectral distribution of a single shot (reference data). We represent the transmission function of the sample as a vector **x**, and thus the vector **T** is equal to the product of the matrix **A** and the vector **x**, namely, **A·x=T**. In the experiment the vector **T** and the matrix **A** are measured, and the equation is solved for the vector **x** using the "total variation minimization by augmented Lagrangian and alternating direction algorithms"

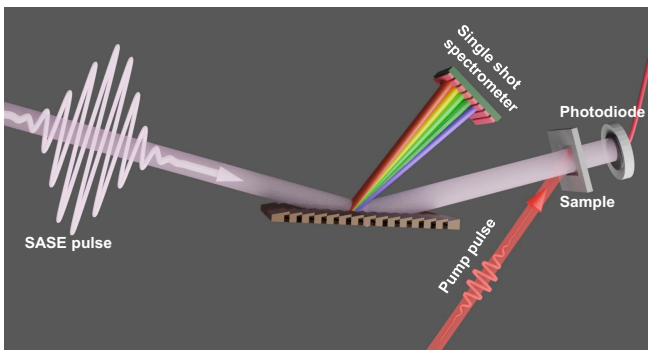

Fig.1. Schematic of the experimental setup. The incident SASE pulse divided into two pulses by a grating present at the PRESTO instrument. The first order diffracted light is measured by a single-shot spectrometer mounted before the sample. The zeroth order irradiates the sample and the transmitted radiation after the sample is measured by a photodiode. See further details in the text.

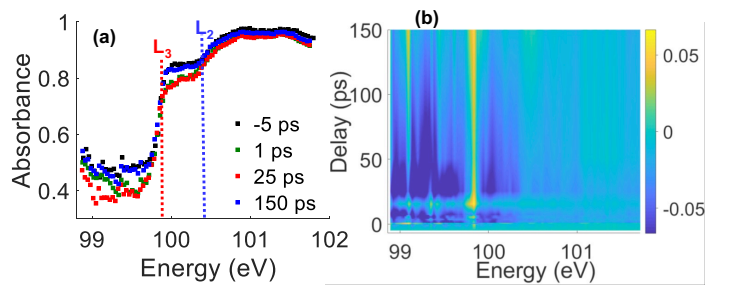

Fig. 2. Photo-induced absorption variation in the silicon membrane (a) XUV absorption spectra at various pump delays. (b) Differential absorption mapping.

(TVAL3) [16, 18].

To demonstrate the photoinduced variation in the L-edge absorption spectrum we plot the spectra at -5 ps (before the arrival of the optical pulse), 1 ps, 25 ps, and 150 ps in Fig. 2a. The largest differences between the photoexcited and the ground state spectrum can be clearly seen at a delay of 1 and 25 ps for photon energies that are slightly above the $L_3$ edge (from 99.8 eV to 100.3 eV) and below the edges.

For a better understanding of the dynamics of the photoinduced effects we numerically calculated the differential absorbance by subtracting absorbance at -5 ps, where the probe arrives before the pump, from the absorbance at a particular delay and plotted the results in Fig. 2b. The color-maps were produced by interpolation with the measured data from -5 to 150 ps.

Inspecting Fig. 2b, we see several different behaviors in the dynamics of the variation of the spectrum. To interpret Fig. 2b, we recall that the $L_3$ and the $L_2$ edges are at about 99.8 eV and 100.3 eV respectively, and that they correspond to excitations from the 2p states (the upper and lower spin states respectively) to the bottom of the conduction band. We first explain the results of Fig. 2b by considering the dynamics of the charge carriers. Since the photon energy of the optical laser is 3.1 eV, it excites electrons from the valence band to the direct valley (the Γ point) of the conduction band. Thus, immediately after the excitation of the electrons, the number of unoccupied states at the Γ point is reduced, and holes are created in the valence band. At delays longer than the optical pulse duration, the charge carriers relax from their excited state, first by electron-electron scattering and at later times (several ps) by electron-phonon scattering [19]. As a result of the relaxation processes, the electrons lose energy and occupy temporarily states at energies lower than the Γ point in the conduction band and inside the gap (i.e. isolated impurity states), finally they eventually relax back to the valence band by Auger recombination after several tens of ps [19]. Thus, we expect the charge carrier dynamics to lead to a reduction in the absorption near 99.8 eV and to an increase in the absorption below 98.7 eV. Since the transition from the 2p level to the Γ point is dipole forbidden [19] and since the temporal resolution in our experiment was ~350 fs, we observed only the slower relaxation processes. The relaxation from the Γ point to the bottom of the conduction band and to the gap states is clearly seen in the form of the reduction in the absorption in the energy range from 98.8 eV to 99.8 eV and from 99.9 eV to 101.7 eV from a delay of 0 ps to a delay of 25 ps and again from a delay of 25 ps to 150 ps. However, there are several interesting features in Fig. 2b that cannot be explained by considering only excitations and relaxations of charge carriers. These are the increased absorption in the narrow area between 99.8 eV and 99.9 eV and the trend of the variation at delays between 1 ps and 7 ps. To focus on these features, we plot in Fig. 3 the delay dependence of the differential absorbance at 99.8 eV and at 100.8 eV, which are the values of the ground state edges and slightly above. The differences between the delay dependencies at the two energies is clear. Above the edges from 0 to 25 ps the absorbance decrease is consistent with the electron relaxation process trend, while the positive variation at 99.8 eV that fluctuates very strongly, and the hump between 1 ps and 7 ps in the curve of the differential absorbance at 100.8 eV, are inconsistent with the relaxation of electrons. The discrepancy between the observations and the explanation based on charge carrier dynamics indicates a possible photo-induced variation of the band structure of the silicon and related structural changes.

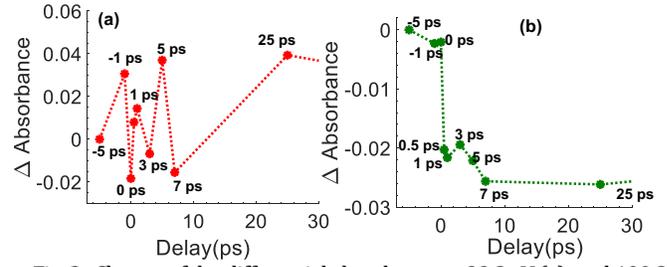

Fig. 3. Change of the differential absorbance at 99.8 eV (a), and 100.8 eV (b) at different delays from -5 to 25 ps.

If these photoinduced variations exist, they should lead to shifts of the edges and to further variations of the absorbance of the sample mainly near the edges. We therefore plot the delay dependence of positions of the edges in Fig. 4.

We estimate the edge position by calculating numerically the derivative of the absorbance spectrum. This derivative peaks at the edges as can be seen in the examples for zero and 25 ps delays shown in Fig. 4a. The $L_3$ edge can be identified at 99.79±0.035 eV and 99.87±0.035 eV at zero and 25 ps delay, respectively. The $L_2$ edge is at 100.30±0.035 eV and 100.40±0.035 eV, at zero and 25 ps delay, respectively. The two peaks below the L edges can be attributed to impurity states in the gap.

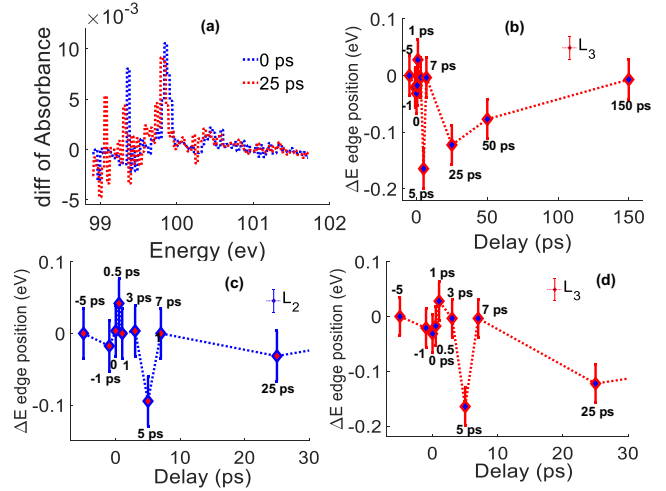

Fig. 4. Variation of the edge position: (a) Derivatives of the absorption spectra of at 0 ps and 25 ps delays. (b) The change of the position of the $L_3$ edge with the delays. (c) and (d) Enlarged view of the change of the position of the $L_2$ and $L_3$ edges respectively from -5 ps to 25 ps.

In figures 4b-4c we can see that the $L_3$ and the $L_2$ edges shift to lower photon energies from 3 ps to 5 ps both. From 5 to 7 ps both edges returned to their original positions. Again from 7 ps to 25 ps both the edges exhibit negative shifts and then they return to their original position at 150 ps. The largest shift of the $L_3$ edge is -0.08 eV and the largest shift of the $L_2$ edge is -0.10 eV. These results indicate that a few ps after the excitations the conduction band is shifted toward the valance band. This can explain the positive variation of the absorption just below the edge, which we see in Fig. 2b, and suggests that the variation is due to the shift of the conduction band and the corresponding change in the density of states near the bottom of the band. The faster shift can be attributed to electron-phonon interaction and the shift at longer delays to the change in the structure of the silicon that leads to the variation in the band gap. The non-monotonic behavior has to be investigated in detail and suggests that several processes can impact the variation of the band gap, for example, temperature and pressure due to the charge carriers. Since the spectral resolution in our experiment was 35 meV, the small positive shifts we see in Fig. 4 are smaller than our experimental precision.

Before concluding we compare our results with pertinent work on transient XUV spectroscopy in silicon. Leone and colleagues [19] reported a detailed study of the transient photoinduced variation of the absorption spectrum by using a high harmonic generation source as the probe and with optical pumps at several wavelengths. The time resolution in their experiment was higher than ours, thus they could monitor the dynamics on the 100-fs scale. The intensity of the optical laser they used was about an order of magnitude weaker than the intensity we used, and their spectral resolution (500 meV) was lower than ours. Most of our results agree with that work and with the theoretical models they suggested. The differences are due to the higher intensity in the present work, which led to more pronounced variations in the band structure. Since we had a better spectral resolution, we observed several structures they did not observe (the edge shifts, in particular), but they all agree with the theory.

Beye et al. [20] reported a photoinduced phase transition with the optical fluence of 250 mJ/cm$^2$, which is about just 30 times higher than the fluence used here. In their experiment they also used a 400 nm laser, but they measured the XUV fluorescence. The 80 meV for $L_3$ and 200 meV for $L_2$ shift we observed when the optical fluence was 8.5 mJ/cm$^2$ suggests that a photoinduced phase transition can occur only if the fluence dependence of the variation in the band structure is highly nonlinear. This is because the bandgap is about 1.1 eV.

In conclusion, we have demonstrated the applicability of ghost spectroscopy for transient absorption spectroscopy and for the study of photoinduced effects with high photon energy FELs. The experimental setup is simple and can provide high spectral resolution with SASE FELs without additional spectrometers after the sample or monochromatizing the input beam. The method can lead to a novel efficient procedure for transient spectroscopy at high photon energies. It is important to note that the time resolution in our method is limited by the pulse duration of the input pump and probe pulses, while the spectral resolution is limited by the width of the spectral spikes. This trade-off between the time and the spectral resolution, that find its optimum for Fourier transform limited pulses, has to be considered when designing measurements with ghost spectroscopy.

**Data availability.** Data underlying the results presented in this paper are not publicly available at this time but may be obtained from the authors upon reasonable request.

**Funding**.
We acknowledge the Israel Science Foundation (847/21) for support.

## Acknowledgments.
Y.K. gratefully acknowledges the support of the Ministry of Science & Technologies, Israel. We thankfully acknowledge the technical and scientific staff at FERMI for their support during the experiment.